\documentclass{ws-p10x7}

\newcommand{\et}{{\rm E}_{\scriptscriptstyle\rm T}}

\def \Pt {{\rm P}_{\rm T}}
\newcommand{\met}{\mbox{$\protect \raisebox{.3ex}{$\not$}\et \ $}}
\newcommand{\ppbar}{p\bar{p}}

\newcommand{\ttbar}{t\bar{t}}
\newcommand{\bbbar}{b\bar{b}}

\def \mtop {$M_{top} \ $}

\newcommand{\gevc} { {\rm GeV}/c }
\newcommand{\gevcc}{ {\rm GeV}/c^{2} }

\begin{document}

\title{Top Physics from Run 1 and Run 2 Prospects at CDF}

\author{Steven R. Blusk, for the CDF Collaboration}

\address{University of Rochester, Rochester, New York 14628}

\twocolumn[\maketitle\abstract{
We present a summary of top quark physics results from Run 1 at CDF using the Run 1
data sample of 106 pb$^{-1}$. In addition to the precursory measurements of the top quark 
mass and $\ttbar$ cross section, we have performed a number of other analyses which 
test the consistency of the $\ttbar$ data sample with the standard model (SM). Deviations
from SM expectations could provide hints for new physics.
We find that the data are consistent with the SM. While the Run 1 data are
statistically limited, we have shown that the systematic uncertainties are under control 
and thus have layed the groundwork for higher precision tests of the SM in Run 2.
This report describes the Run 1 top quark analyses and expectations and prospects for
top quark measurements in Run 2. }]

\section{Introduction}

	In $\ppbar$ collisions at the Tevatron ($\sqrt{s}$=1.8 TeV), top quark pairs
are produced through the strong interaction with an expected cross section (at NLO) of 
5.1 pb~\cite{ttbar-xs-theory}. 
Single top quarks are also expected to be produced through a t-W-b electroweak vertex with
an expected total cross section of $\approx$1/2 that of $\ttbar$~\cite{singletop-xs-theory}.
Within the SM, the top quark is expected to decay with a lifetime of 
$\approx~10^{-24}$ seconds into a W boson and a b quark. 
$\ttbar$ final states are classified according to the decays modes of the two W bosons. Dilepton
final states consist of events where both W bosons decay to an $e$ or $\mu$ (BR=5\%). Lepton + jets 
final states include events where one of the W bosons decays leptonically ($e$ or $\mu$)
and the other hadronically (BR=30\%). The All-Jets mode includes events in which both W-bosons decay 
hadronically (BR=44\%). 


\vspace{-0.20in}
\section{$\ttbar$ Cross Section}

   Cross section measurements have been made in all three decay 
channels. In the dilepton channel~\cite{dil-xsec}, we observe 9 events with an 
expected background of 2.5$\pm$0.5 events, which leads to a $\ttbar$
cross section of $8.2^{+4.4}_{-3.4}$ pb. In the lepton+$\ge$3 jets
channel, there are 29 (25) events which are SVX (SLT) tagged with expected backgrounds 
of 8.1 (13.2) events, leading to a measurement of $5.7^{+1.9}_{-1.5}$ pb~\cite{ljet-xsec}. 
For the All-Jets mode, we measure $7.6^{+3.5}_{-2.7}$ pb~\cite{allhad-mass-xsec}. Results 
from all three channels are combined to obtain a $\ttbar$ cross section of 
$6.5^{+1.7}_{-1.4}$ pb~\cite{ptohos} which is within one standard deviation from 
the theoretical prediction.

\vspace{-0.15in}
\section{Top Quark Mass}

   The most precise measurements in the top quark sector thus far have been in the mass.
In the dilepton channel, we use a weighting technique which compares the
observed \met in each event to the expected value as a function of the assumed top mass.
Using a likelihood technique we extract a top mass of \mtop=167.4$\pm$11.4~$\gevcc$~\cite{dil-mass}.
In the lepton+$\ge$4 jets events, we perform a 2C fit of the final state particles to the decay 
chain, which results in a measured top mass of $176.1\pm7.4~\gevcc$~\cite{ljet-mass}.
Full reconstruction of events in the All-Jets mode is also performed from which we 
measure \mtop=186.0$\pm 11.5~\gevcc$~\cite{allhad-mass-xsec}. 
The result from combining all three measurements is
$176.1\pm 6.6~\gevcc$, roughly 35 times the mass of the next heaviest quark!


\vspace{-0.15in}
\section{The $\ttbar$ Invariant Mass ($M_{\ttbar}$)}

\begin{figure}
\epsfxsize160pt
\vspace{-0.25in}
\figurebox{120pt}{160pt}{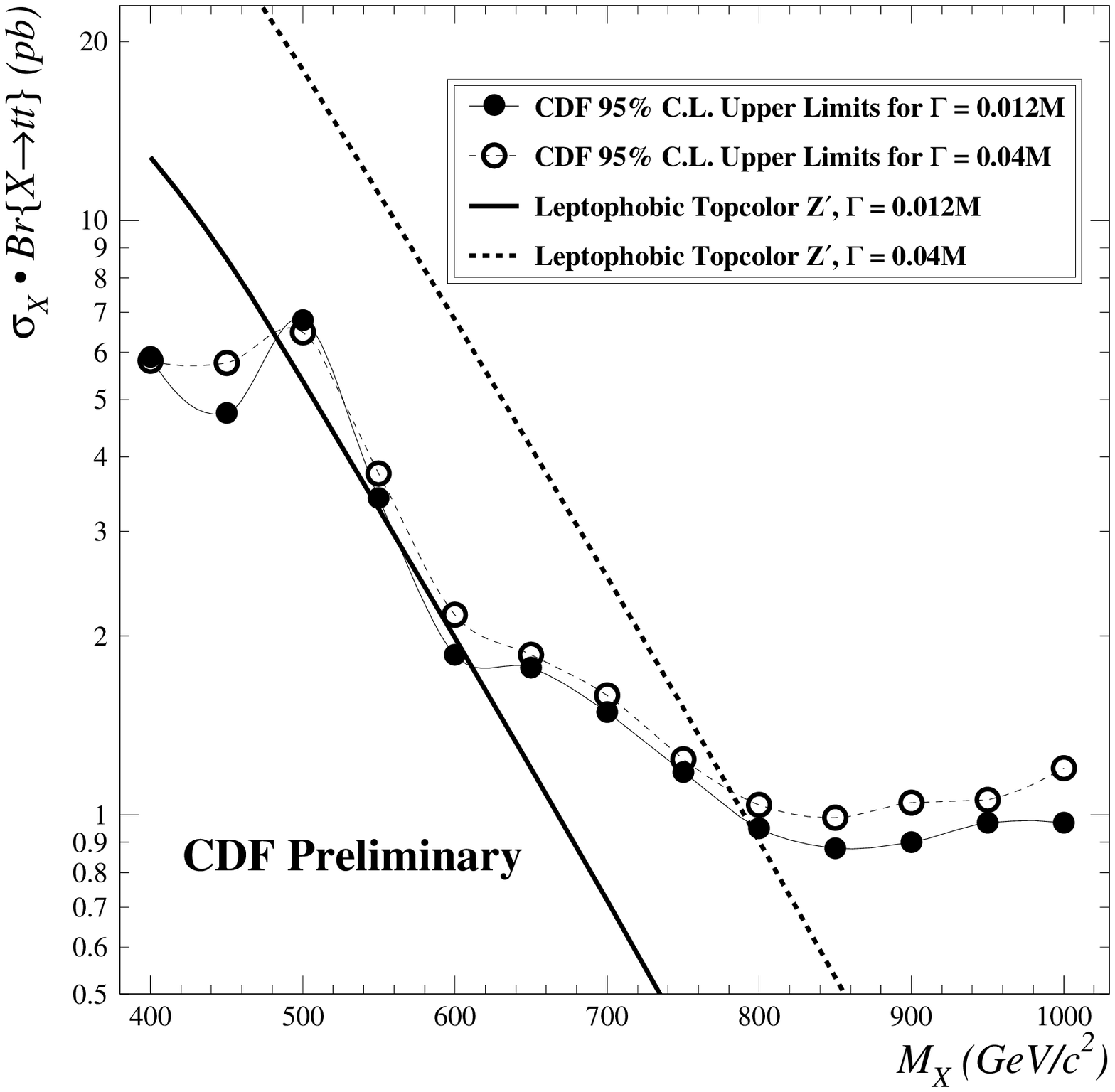}
\vspace{-0.4in}
\caption{The 95\% CL limits on $\sigma_X\dot BR(X\to\ttbar)$ as a function of
the mass of $X$ for two different values of the full width of the heavy object. The data are 
compared to the prediction for a leptophobic topcolor $Z^{\prime}$ for full widths of 0.012 
$M_{Z^{\prime}}$ and 0.04 $M_{Z^{\prime}}$.}
\label{fig:mttbar-limits}
\end{figure}

The $M_{\ttbar}$ analysis~\cite{mttbar} proceeds in a similar way to the top mass analysis.
To improve the resolution on the four momenta of the final state particles (and thus $M_{\ttbar}$),
we constrain the top quark mass to 175 $\gevcc$ in the fit. We also require when we remove
this constraint that the fitted top quark mass lie in the range from 150-200 $\gevcc$. 
The data do not show an excess above the SM prediction, and we therefore present
limits on the cross section times branching ratio (see Fig.~\ref{fig:mttbar-limits}).
At the 95\% confidence level (CL), 
the data rule out a topcolor $Z^{\prime}$ with mass less than 480 (780) $\gevcc$ and natural width
equal to 0.012 (0.04) $M_{Z^{\prime}}$.


\vspace{-0.15in}
\section{Top $\Pt$}

Like the $M_{\ttbar}$ analysis, we use $l$+4 jet data and constrain the top quarks mass to 
175 $\gevcc$. Because of the strong correlation between the top and antitop quarks' $\Pt$,
we use only the hadronically decaying top quark. We measure
the fraction of top quarks produced in four bins of true $\Pt$: 0-75 $\gevcc$, 
75-150 $\gevcc$, 150-225 $\gevcc$, and 225-300 $\gevcc$. First, we determine initial 
response functions which give the distribution of reconstructed $\Pt$ in each of the four 
true $\Pt$ bins. The data are then fit to a combination of the
four Monte Carlo (MC) reconstructed $\Pt$ distributions using an iterative procedure to minimize 
the sensitivity of the final result to the initial assumptions of the true top $\Pt$ distribution.
Within the limited statistics, the data are consistent with  
SM expectations. We measure the 95\% CL limit for the fraction of top 
quarks with true $\Pt$ larger than 225 $\gevc$ to be 0.114.



\vspace{-0.15in}
\section{W Helicity in Top Decays}

  The V-A structure of the t-W-b vertex results in a specific prediction for the W
polarization in top decays. At tree level, we expect the fraction of longitudinal $W$
bosons, $F_0$, to be 70.1$\pm$1.6\%. The $\Pt$ spectrum of the leading lepton is sensitive to
the W polarization. Using MC distributions of longitudinal and left-handed W's,
we fit the data to extract the fraction $F_0$. Using both the lepton+jets and dilepton
data samples, we measure $F_0=91\pm 37(stat)\pm 13$\%~\cite{w-hel}. 

\vspace{-0.15in}
\section{Rare Decays}

    The FCNC decays $t\to Zq$ and $t\to\gamma q$ are strongly suppressed in the S.M.
at the level of $\sim$10$^{-12}$, and therefore an observation of such events is a
signature of new physics. We have performed searches for these decays~\cite{rare-decays}
and find one event in each channel, consistent with background expectations. We therefore
set 95\% CL limits of 33\% and 3.2\% respectively for these two FCNC decays.

\vspace{-0.15in}
\section{Single Top Production}

   We have searched for single top in the lepton+jets data. One analysis searches for events in
both the W-gluon fusion and the s-channel $W^*$ processes. We select W+1,2,3 jet events which
have a SVX b-tag and a top invariant mass, $M_{l\nu b}$ in the range 140 to 210 $\gevcc$. 
We observe 65 events with an expected background of 62.5$\pm$11.5 events. We expect to 4.3 signal
events. Fitting the $H_T=\sum \et (lepton, \met, jets)$ distribution in data to MC 
signal and background distributions, we extract a cross section limit of 13.5 pb at 95\% CL. 
A second analysis which looks just for the W-gluon
fusion process selects W+2 jet events with an SVX tag and the same cut on $M_{l\nu b}$. An 
interesting and exploitable feature of these events is that, unlike the backgrounds, the 
product of the leading lepton's charge ($Q$) and the pseudorapidity of the untagged jet 
($\eta$) peaks at positive $Q\times\eta$. We observe 15 events with an expected background 
of 12.9$\pm$2.1 events (we expect 1.2$\pm$0.3 signal events). From a fit of the 
$Q\times\eta$ distribution in data to MC signal and background distributions, we extract 
a 95\% CL limit of 15.4 pb.

\vspace{-0.2in}
\section{Run 2 Expectations}
\vspace{-0.1in}

   Run 2 will provide $\approx$40-50 times more $\ttbar$ events than Run 1. In addition to a 
large reduction in statistical uncertainties, systematic uncertainties such as the jet energy 
scale and MC modelling will also be reduced. For example, the large 
sample of $Z\to\bbbar$ events can be used to check the $b-jet$ energy scale. The invariant mass
of the two untagged jets in double SVX tagged W+4 jet events can be used to check the light quark 
jet energy scale. A comparison of extra jets in a high purity top sample can be used to put 
constraints on gluon radiation in the MC simulation. Moreover, we expect to undertake new physics
analyses in Run 2, such as studying the spin correlations in $\ttbar$ events.
Given the size of the Run 2 data sample, we have made projections for the precision we can expect
for a variety of measurements. Some of these projections are given in Table~\ref{tab:run2proj}.
Run 2 and a future Run 3 will clearly provide very rich top samples with which to probe the SM and 
beyond.

\begin{table}
\caption{Projections for the expected precision for measurements with an integrated luminosity 
of 2 $fb^{-1}$.}\label{tab:run2proj}
\begin{tabular}{|c|c|c|c|} 
\hline 
Measurement &  Precision \\
\hline
\mtop 				& 1.5\%  \\
$\ttbar$ cross section  	& 9\%    \\
Single top cross section	& 24\%   \\
$V_{tb}$ (from Single top)	& 13\%   \\
$F_0$				& 5.5\%  \\
$\sigma*BR(X\to\ttbar )$	& 0.1 pb at 1 TeV \\
$ BR(t\to \gamma c)$		& $<$2.8x10$^{-3}$ \\
$ BR(t\to Zc)$			& $<$1.3x10$^{-2}$ \\
$ BR(t\to Hb)$			& $<12$\% \\
\hline
\end{tabular}
\end{table}

\vspace{-0.15in}
\section*{Acknowledgments}
\vspace{-0.1in}
   We thank the Fermilab staff and our CDF collaborators for their vital 
contributions to these physics analyses.

\end{document}